# Model Predictive Path Integral Control for Roll-to-Roll Manufacturing

Christopher Martin*, Apurva Patil*, Wei Li*, Takashi Tanaka**, Dongmei Chen*

*The University of Texas at Austin, Austin, TX 78712 USA (email: cbmartin129@utexas.edu, apurvapatil@utexas.edu, weiwli@austin.utexas.edu, dmchen@me.utexas.edu)
**Purdue University, West Lafayette, IN 47907 USA (email: tanaka16@purdue.edu)

**Abstract**: Roll-to-roll (R2R) manufacturing is a continuous processing technology essential for scalable production of thin-film materials and printed electronics, but precise control remains challenging due to subsystem interactions, nonlinearities, and process disturbances. This paper proposes a Model Predictive Path Integral (MPPI) control formulation for R2R systems, leveraging a GPU-based Monte-Carlo sampling approach to efficiently approximate optimal controls online. Crucially, MPPI easily handles non-differentiable cost functions, enabling the incorporation of complex performance criteria relevant to advanced manufacturing processes. A case study is presented that demonstrates that MPPI significantly improves tension regulation performance compared to conventional model predictive control (MPC), highlighting its suitability for real-time control in advanced manufacturing.

## 1. INTRODUCTION

Roll-to-Roll (R2R) manufacturing is a high-speed, continuous processing method that has historically been used for substrate-based products such as papers, sheet metals, and fabrics (Sheats, 2002; Martin et al., 2025a). Its advantages include efficiency and scalability, making it a promising approach for advanced applications such as printed electronics (Pastorelli *et al.*, 2016) and thin-film materials (Hong *et al.*, 2022). However, maintaining consistent product quality remains a challenge compared to traditional batch processes. Accurate web tension control is particularly crucial, as tension variations lead to wrinkles, tears, slack, and registration errors (Raul and Pagilla, 2015; Chen *et al.*, 2023). These challenges necessitate advancements in R2R control to meet the growing demands for precision in complex manufacturing applications.

Over the past several decades R2R control has evolved from decentralized proportional-integral (PI) controllers to more sophisticated model-based approaches. PI controllers have been widely used due to their simplicity, but they struggle with subsystem interactions and process variations (Young and Reid, 1993; Raul and Pagilla, 2015). To account for subsystem interactions, disturbances, and uncertainties, centralized state-space and robust control techniques have been proposed (Xu, 2009; Martin et al., 2025b). More recently, model predictive control (MPC) methods have been proposed to rigorously account for input constraints, future disturbances, and register error online (Shah *et al.*, 2022; Chen *et al.*, 2023).

These MPC methods are effective, but they rely on solving an optimal control quadratic program (QP) over a prediction horizon at each timestep. Without the use of custom solvers, solving these QPs can have a complexity of $O(N^3H^3)$ in the worst case, where $N$ represents the number of actuated rollers in an R2R line and $H$ is the prediction horizon (Wang and Boyd, 2010). This computation complexity can be a barrier for online control of large R2R lines with hundreds of states (Martin, et al., 2025a). Additionally, these optimization-based MPC methods can only handle quadratic, or at minimum continuous, cost functions. This is a significant drawback, as more intricate non-differentiable cost functions can be useful for advanced manufacturing. Thus, a method is needed for large R2R manufacturing lines that can calculate the optimal control online with respect to diverse cost functions that reflect the complexity of advanced manufacturing processes.

A promising solution to this need is model predictive path integral (MPPI) control. MPPI is a stochastic optimal control framework that allows the controller to compute the optimal control input on the fly using online (real-time) Monte-Carlo simulations (Kappen, 2005; Williams *et al.*, 2016). The path integral control framework was pioneered by Kappen (Kappen, 2005) and is gaining popularity in control applications (Janson, Schmerling and Pavone, 2017; Patil *et al.*, 2023). This growing interest is due to its generality and scalability. Unlike many existing control synthesis methods, it can directly deal with stochasticity and nonlinearity; and unlike stochastic dynamic programming, it can evaluate control input without solving a high-dimensional optimal control problem (Patil *et al.*, 2022).

The path integral approach is rooted in the principles of statistical physics. Based on the Feynman-Kac theorem (Kappen, 2005), the path integral method transforms the optimal control problem into an expectation over noisy system trajectories, enabling efficient policy computation through Monte Carlo sampling. According to the strong law of large numbers (Durrett, 2019), as the number of Monte Carlo samples increases, the policy generated by the path integral method converges to optimality. It has also been demonstrated that Monte-Carlo simulations for path integral control can be massively parallelizable and thus can leverage the full potential of GPU resources that are now standard in many control platforms (Williams, Aldrich and Theodorou, 2017).

Due to its purely Monte-Carlo-driven nature, MPPI is particularly well-suited for high-dimensional, nonlinear, and stochastic problems, including R2R manufacturing. MPPI's Monte-Carlo style does not depend on approximate linear models to compute control inputs, granting it greater flexibility and robustness. Furthermore, by directly incorporating system stochasticity, the Monte-Carlo approach ensures that control policies are optimized not only for expected outcomes but also for effectively handling variability in system behavior. Additionally, MPPI can seamlessly manage non-differentiable cost functions by evaluating them explicitly along each sample path. These benefits contrast with deterministic optimization-based MPC methods that have been applied to R2R systems in the past, which struggle with incorporating stochasticity and require differentiable cost functions and system models (Shah *et al.*, 2022; Chen *et al.*, 2023).

Another important consideration is that MPPI performs best in systems where stochasticity enters the dynamics via the control channel. Fortuitously, disturbances in R2R processing lines primarily appear precisely in those signals over which control authority is exercised, aligning naturally with MPPI's structure. Specifically, the control inputs in R2R systems are typically the roller motor torques, and most critical disturbances—such as motor friction or gear backlash—also manifest as torques acting directly on the rollers (Dwivedula and Pagilla, 2013; Kang *et al.*, 2021). Additionally, due to the distributed structure of R2R lines, the computational cost of MPPI's Monte Carlo rollouts scales linearly with the size of the system, unlike the super-linear growth characteristic of typical QP-based MPC (Wang and Boyd, 2010; Shah *et al.*, 2022; Chen *et al.*, 2023). Thus, MPPI is an excellent candidate for achieving high-performance control in R2R systems.

This paper presents an MPPI formulation for R2R control. The method uses a Monte-Carlo approach to approximate the optimal control signal instead of solving an optimization problem online. This approach means that the control can be cheaply computed online for large R2R lines using GPUs; and since the full nonlinear model can be used in the Monte-Carlo simulations, no linear approximations of the system are necessary to synthesize control. Finally, and most importantly for this application, the MPPI approach can easily handle non-quadratic, or even non-differentiable, cost functions, enabling the full complexity of advanced manufacturing processes to be accounted for in the control policy.

The paper is structured as follows. Section 2 presents the stochastic dynamic model of an R2R line, while Section 3 formulates the R2R MPPI control problem. Section 4 develops an MPPI controller for R2R systems. Section 5 provides a case study analyzing the proposed method on a standard R2R line, and Section 6 concludes the study.

## 2. DYNAMIC MODEL

A simplified schematic of an R2R line is given in Figure 1, and the nomenclature associated with this R2R system is listed in Table 1. The assumptions that give rise to the model presented in this section are as follows (Shelton, 1986). 1) Passive rollers do not contribute significantly to the system dynamics, so only actuated (motorized) rollers are considered (Martin et al., 2025a). 2) Tension is equivalent everywhere between two adjacent rollers. 3) Friction is high enough that there is no web slippage over the rollers, so $v_i = \omega_i R_i$.

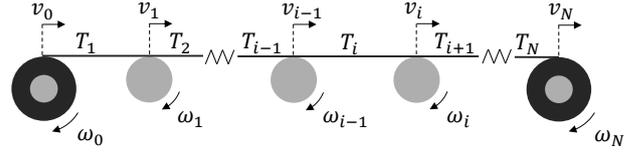

Figure 1: Schematic of a simplified R2R line.

Table 1: R2R nomenclature

| Parameter/signal | Variable |
|---|---|
| Modulus of web | $E$ (N) |
| Cross-sectional area of web | $A$ (m$^2$) |
| Radius of roller $i$ | $R_i$ (m) |
| Inertia of roller $i$ | $J_i$ (kg-m$^2$) |
| Friction coefficient motor $i$ | $f_i$ (N-m-s-rad$^{-1}$) |
| Length of web section $i$ | $L_i$ (m) |
| Disturbance coefficient motor $i$ | $b_i$ (s-m$^{-1}$-kg$^{-1}$) |
| Tension in web section $i$ | $T_i$ (N) |
| Linear velocity over roller $i$ | $v_i$ (m-s$^{-1}$) |
| Unwinding velocity | $v_0$ (m-s$^{-1}$) |
| Rotational velocity of roller $i$ | $\omega_i$ (rad-s$^{-1}$) |
| Motor torque applied to motor $i$ | $u_i$ (N-m) |

The web tension dynamics, which can be determined using conservation of mass (Shelton, 1986), are as follows:

$$dT_i = \left(\frac{AE}{L_i}(v_i - v_{i-1}) + \frac{1}{L_i}(T_{i-1}v_{i-1} - T_i v_i)\right)dt \quad (1)$$

Additionally, the roller velocity dynamics, which can be determined using a torque balance, are as follows:

$$dv_i = \left(\frac{R_i^2}{J_i}(T_{i+1} - T_i) - \frac{f_i}{J_i}v_i + \frac{R_i}{J_i}u_i\right)dt + b_i dw_i \quad (2)$$

where $T_0 = T_{N+1} = 0$ and $dw_i$ is a Brownian disturbance related to motor friction, backlash from gears, or other disturbances (Dwivedula and Pagilla, 2013; Kang *et al.*, 2021). Time dependence is omitted for brevity.

## 3. CONTROL PROBLEM FORMULATION

Equations (1) and (2) give the stochastic R2R dynamics in continuous time. For implementation on a digital computer, it is necessary to discretize these equations as follows.

$$T_i(t+1) = T_i(t) + \left(\frac{AE}{L_i}(v_i(t) - v_{i-1}(t)) + \frac{1}{L_i}(T_{i-1}(t)v_{i-1}(t) - T_i(t)v_i(t))\right)\Delta t \quad (3)$$

$$v_i(t+1) = v_i(t) + \left(\frac{R_i^2}{J_i}(T_{i+1}(t) - T_i(t)) - \frac{f_i}{J_i}v_i(t) + \frac{R_i}{J_i}u_i(t) + \frac{b_i}{\sqrt{\Delta t}}\epsilon_i(t)\right)\Delta t \quad (4)$$

where $\epsilon_i(t) \in \mathcal{N}(0,1)$ and $\Delta t$ is a fixed sampling rate or step size. The $\epsilon_i(t)$ term is the discretized version of the continuous Brownian disturbance term from Eq. (2) (Williams, Aldrich and Theodorou, 2017). These dynamics can be concatenated into the following discrete state-space representation:

$$x_{t+1} = x_t + \left(f(x_t, t) + Gu_t + \frac{B}{\sqrt{\Delta t}}\epsilon_t\right)\Delta t \quad (5)$$

where $x_t = [T_1(t), T_2(t), ..., T_N(t), v_1(t), v_2(t), ..., v_N(t)]^T$, $u_t = [u_1(t), u_2(t), ..., u_N(t)]^T$, and $\epsilon_t \sim \mathcal{N}(0, I_{N \times N})$. Note that Eq. (5) can be partitioned as follows:

$$f(x_t, t) = \begin{pmatrix} f_a(x_t, t) \\ f_c(x_t, t) \end{pmatrix}, x_t = \begin{pmatrix} x_t^a \\ x_t^c \end{pmatrix}, \quad G = \begin{pmatrix} 0_{N \times N} \\ G_c \end{pmatrix}, B = \begin{pmatrix} 0_{N \times N} \\ B_c \end{pmatrix} \quad (6)$$

where $f_a(x_t, t): \mathbb{R}^{2N} \times \mathbb{R}^+ \to \mathbb{R}^N$, $f_c(x_t, t): \mathbb{R}^{2N} \times \mathbb{R}^+ \to \mathbb{R}^N$, $x_t^a = [T_1(t), T_2(t), ..., T_N(t)]^T \in \mathbb{R}^N$, $x_t^c = [v_1(t), v_2(t), ..., v_N(t)]^T \in \mathbb{R}^N$, $G_c \in \mathbb{R}^{N \times N}$, and $B_c \in \mathbb{R}^{N \times N}$. Thus, there is direct control authority over a state if and only if there is also a disturbance acting on that state. This condition is a prerequisite for MPPI control (Kappen, 2005; Williams, Aldrich and Theodorou, 2017).

Given this dynamic structure, the question motivating this paper is how to optimally control the R2R system with respect to a cost function of the following form:

$$J = \min_u \mathbb{E} \left[ \phi(x_H, H) + \sum_{t=1}^{H-1} q(x_t, t) + \frac{1}{2} u_t^T R u_t \right] \quad (7)$$

where $q(x_t, t)$ is the stage cost for the state (in the R2R context, $q(x_t, t)$ will typically penalize deviation from a desired tension and velocity setpoint), $\phi(x_H, H)$ is the terminal cost, $H$ is the receding optimization horizon, the expectation is taken with respect to the discrete dynamics (5), and:

$$R = \lambda G_c^T B_c^{-T} B_c^{-1} G_c \quad (8)$$

meaning that on channels where the disturbance magnitude is high, control is cheap. This condition is also a prerequisite for MPPI control (Kappen, 2005; Williams, Aldrich and Theodorou, 2017). Additionally, note that the cost function is general with respect to the state, with no constraints on differentiability. Managing such a cost function is challenging for typical optimization-based controllers such as iterative linear quadratic gaussian control (iLQG) (Todorov and Li, 2005) or sequential quadratic programming-based MPC (Gros *et al.*, 2020), as they rely on local quadratic approximations of the cost function, which is impractical if the function is non-differentiable. Thus, the R2R system structure in Eqs. (5) and (6) and desired cost function (7) motivate the use of MPPI control. This problem formulation is illustrated in Figure 2.

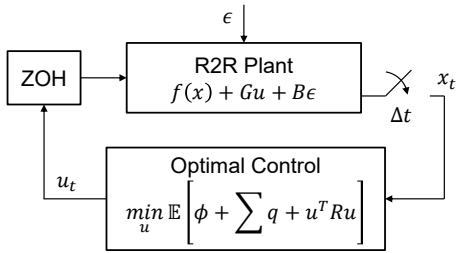

Figure 2: R2R MPPI control problem formulation.

## 4. CONTROL DESIGN

### 4.1 Path integral control over a finite horizon

This section presents the proposed method to solve the receding horizon optimal control problem summarized by Eq. (7) and Figure 2. The method has a path integral structure, where the optimal control is found using a Monte-Carlo approach with importance sampling (Kappen, 2005; Williams, Aldrich and Theodorou, 2017). Importance sampling allows precise tuning of exploration around a baseline control policy, rather than the open-loop dynamics, enabling drastic improvements in computational efficiency as only relevant portions of the policy space are explored.

To start, note that the discrete dynamics formulation Eq. (5) can be arranged in the following manner:

$$x_{t+1} = x_t + \left( f(x_t, t) + G \left( u_t + \frac{D}{\sqrt{\Delta t}} \epsilon_t \right) \right) \Delta t \quad (9)$$

where $D = G_c^{-1} B_c$. Though Eq. (9) represents the true stochastic system behavior, it would be inefficient to find a control policy sampling from these dynamics as most of the simulated behavior would be far from optimal. Instead, we use an importance sampling approach (Kappen, 2005; Williams, Aldrich and Theodorou, 2017). To solve the optimal control problem (7), $K$ Monte-Carlo trials, or "rollouts," of the following adjusted system are performed:

$$\begin{aligned} x_{t+1} &= x_t + f(x_t, t) \Delta t + G \left( \bar{u}_t + \sqrt{\nu} \frac{D}{\sqrt{\Delta t}} \epsilon_t \right) \Delta t \\ &= x_t + f(x_t, t) \Delta t + G(\bar{u}_t + \sqrt{\nu} \delta u_t) \Delta t \end{aligned} \quad (10)$$

where $\bar{u}_t$ is a nominal control, $\nu$ is a user-chosen parameter that tunes exploration, and $\delta u_t = \frac{D}{\sqrt{\Delta t}} \epsilon_t$. Thus, $\nu$ and $\bar{u}_t$ are parameters that can be tuned to optimize exploration in the policy space, while $\delta u_t$ is the natural variance of the system. Using the rollouts from this adjusted system, the optimal control with respect to the true system at time $t$ can be calculated as follows:

$$u_t^* = \bar{u}_t + \frac{\sum_{k=1}^{K} \exp(-S_k/\lambda) \delta u_t^k}{\sum_{k=1}^{K} \exp(-S_k/\lambda)} \quad (11)$$

where $\delta u_1^k, \delta u_2^k, ..., \delta u_{H-1}^k$ is the disturbance for the $k^{\text{th}}$ Monte Carlo rollout, $\lambda$ is the "temperature" parameter that tunes how aggressively the policy follows the best rollouts, and

$$S_k = \phi(x_H, H) + \sum_{t=1}^{H-1} \tilde{q}(x_t, \bar{u}_t, \delta u_t^k, t) \quad (12)$$

where,

$$\tilde{q}(x, \bar{u}, \delta u, t) = q(x, t) + \frac{1 - \nu^{-1}}{2} \delta u^T R \delta u + \bar{u}^T R \delta u + \frac{1}{2} \bar{u}^T R \bar{u} \quad (13)$$

Thus, the finite horizon optimal control problem (7) can be approximately solved through performing many Monte-Carlo simulations. Critically, these rollouts can be performed in parallel with a modern GPU, enabling real-time implementation (Williams, Aldrich and Theodorou, 2017).

### 4.2 Model predictive path integral control

This finite-horizon path integral optimal control solution can be integrated into an MPC framework (Williams *et al.*, 2016; Williams, Aldrich and Theodorou, 2017). Specifically, at each timestep $t$, the path integral Monte Carlo method summarized by Eqs. (11)-(13) is used to find a control sequence $u_{1|t}, u_{2|t}, ..., u_{H-1|t}$ that approximately solves the finite horizon control problem (7). $u_{1|t}$ is then used at time step $t$, and the process is repeated at the next time step. This control approach, called model predictive path integral (MPPI) control, is summarized in Algorithm 1.

---

**Algorithm 1: Model Predictive Path Integral Control**

**Given:** $K$: Number of samples; $H$: Horizon length (time steps); $u_{1|0}, u_{2|0}, \ldots, u_{H-1|0}$: Initial control sequence; $\Delta t, f, G, D, \nu$: System/sampling dynamics; $q, \lambda$: Cost parameters

**while** task not completed, **do**
  Measure $x_t$
  **for** $k \leftarrow 1$ to $K$, **do**
    $x_{1|t}^k \leftarrow x_t$
    Generate $\delta u_{1|t}^k, \delta u_{2|t}^k, \ldots \delta u_{H-1|t}^k$
    **for** $i \leftarrow 1$ to $H-1$, **do**
      $x_{i+1|t}^k = x_{i|t}^k + \left(f(x_{i|t}^k, t) + G(u_{i|t} + \sqrt{\nu}\delta u_{i|t}^k)\right)\Delta t$
      $S_k\mathrel{+}= \tilde{q}(x_{i|t}^k, u_{i|t}, \delta u_{i|t}^k, i)$
    $S_k\mathrel{+}= \phi(x_{H|t}^k, H)$
  **for** $i \leftarrow 1$ to $H-1$, **do**
    $u_{i|t}\mathrel{+}= \left(\sum_{k=1}^{K}\exp(-S_k/\lambda)\,\delta u_{i|t}^k\right)/\left(\sum_{k=1}^{K}\exp(-S_k/\lambda)\right)$
  Send to actuators $u_{1|t}$
  **for** $i \leftarrow 1$ to $H-2$, **do**
    $u_{i|t+1} = u_{i+1|t}$
  $u_{H-1|t+1} = Initialize(u_{H-1|t})$
  $t\mathrel{+}= 1$

---

## 5. CASE STUDY

In this case study we apply MPPI control to an industrial R2R line. To demonstrate the flexibility of the MPPI approach, we use two different cost functions, summarized as follows.

$$q(x_t, t) = \|x_t - x_t^r\|_Q \quad (14.A)$$

$$q(x_t, t) = \|x_t - x_t^r\|_Q + \sum_{i=1}^{N} q_{L_1}|T_i(t) - T_i^r(t)| \quad (14.B)$$

where $\|z\|_M = z^T M z$, $q_{L_1}$ is a term that penalizes the $L_1$-norm of the tension deviation, $x_t^r$ represents the reference, or desired, state trajectory, and $T_i^r(t)$ represents the reference tension trajectory for web section $i$. The first cost function is differentiable and quadratic and is thus compatible with optimization-based MPC approaches (Shah *et al.*, 2022; Chen *et al.*, 2023). The second cost function involves a non-differentiable $L_1$ term, which would be difficult to implement using a standard optimization-based MPC. In addition to these two MPPI controllers, we also implement a standard linear MPC (LMPC) with a quadratic cost function as a baseline (Shah *et al.*, 2022; Chen *et al.*, 2023). At each timestep, this controller uses a linear model of the system to find the optimal control with respect to the quadratic cost over the time horizon.

### 5.1 Simulation setup

This case study considers a standard R2R line, as illustrated in Figure 1. All states are assumed to be measurable through encoders and load cells (Martin et al., 2025a). The line has six web sections, where, for simplicity, the physical parameters for each web section and roller are the same. These parameters, chosen to resemble a typical R2R line (Zhao *et al.*, 2022; Chen *et al.*, 2023), are listed in Table 2.

The hyper-parameters for the proposed and baseline controllers were chosen as follows. First, the quadratic cost on the state error, $Q$, was the same for all three controllers and was tuned using Bryson's rule (Bryson, 1993), where $Q_{i,i} \sim x_i^{-2}$. Additional weight was given to the cells of $Q$ associated with the web tensions. The quadratic control penalty, $R$, for the LMPC was also tuned using Bryson's rule, and it was scaled so that the control effort of the LMPC had the same maximum amplitude as the MPPI controllers. The prediction horizon, $H$, was 9 time steps for all three controllers. For the MPPI controllers, $\lambda = 1$ and $\nu = 1$. For the MPPI controller with an $L_1$ term, $q_{L_1} = 100 \times Q_{1,1}$. The number of Monte Carlo rollouts was set to $K = 3\times10^4$, which corresponded to the maximum number of rollouts achievable on our machine within the sampling time, $\Delta t = 0.01$ s.

Table 2: Physical parameters of the case study

| Parameter | Numerical value |
|---|---|
| Modulus ($E$) | 200 MPa |
| Cross-sectional area ($A$) | $1.2 \times 10^{-5}$ |
| Roller radius ($R$) | 0.04 m |
| Roller inertia ($J$) | 0.95 kg-m$^2$ |
| Motor friction ($f$) | 10 N-m-s-rad$^{-1}$ |
| Web section length ($L$) | 1.0 m |
| Disturbance coefficient ($b$) | $1.0 \times 10^{-2}$ (s-m$^{-1}$-kg$^{-1}$) |

The rest of this section will investigate two different operating cases. The first is the system response to a step increase in the reference tension of one of the web sections. The purpose of this operating case is to evaluate the controllers' ability to independently control the tension in each web. The second operating case will investigate the system response to a speed-up in the unwinding velocity. This "speed-up" phase is a common subject of investigation in R2R systems, as maintaining proper tension control as the roller speeds change is challenging (Chen *et al.*, 2019; Zhang, Tan and Chen, 2024).

### 5.2 Tension control

In this test case, webs 1, 2, 4, 5, and 6 will be set at 28 N, 36 N, 40 N, 24 N, and 32 N, respectively. Web 3 will start at 20 N and then jump to 44 N. The unwinding velocity is set to 0.01 m-s$^{-1}$. To ensure that the reference operating point is in equilibrium, the reference roller velocities were set to:

$$v_i^r(t) = \frac{EA - T_{i-1}(t)}{EA - T_i(t)} v_{i-1}^r(t), v_0^r = v_0 \quad (15)$$

The results of this test case are shown in Figure 3. Also, the maximum tension deviations and convergence times for each of the six web sections are listed in Table 3 for each of the three controllers. The convergence time for each web section is defined as the time it takes to return to within 0.5 N of the corresponding reference tension value, and the maximum tension deviations are also relative to these reference values.

Table 3: Tension control results statistics.

| Web section | | 1 | 2 | 3 | 4 | 5 | 6 |
|---|---|---|---|---|---|---|---|
| Convergence time (s) | LMPC | 0.65 | 0.77 | 0.61 | 0.54 | 0.50 | 0.31 |
| | MPPI w/o L1 | 0.82 | 0.95 | 0.90 | 0.95 | 0.55 | 0.26 |
| | MPPI w/ L1 | 0.33 | 0.29 | 0.27 | 0.21 | 0 | 0 |
| Maximum Tension Deviation (%) | LMPC | 12.7 | 14.3 | 0.41 | 11.8 | 8.04 | 2.59 |
| | MPPI w/o L1 | 11.4 | 13.5 | 0.80 | 11.1 | 9.38 | 3.22 |
| | MPPI w/ L1 | 9.18 | 11.0 | 0.18 | 6.63 | 0.75 | 0.53 |

As can be seen in the figure and the table, the MPPI controller without the $L_1$ penalty term achieves approximately the same performance as the LMPC. This is to be expected since they have the same cost function. In contrast, the MPPI controller with the $L_1$ penalty term achieves significantly better performance compared with the LMPC. Specifically, the MPPI controller with the $L_1$ term had an average convergence time of 0.18 s and an average maximum tension deviation of 4.71%, while the LMPC had an average convergence time of 0.56 s and an average maximum tension deviation of 8.30%. This contrast illustrates the utility of the non-differentiable $L_1$ term, emphasizing the importance of the MPPI controller's ability to include cost terms that are difficult to implement in QP optimization-based schemes

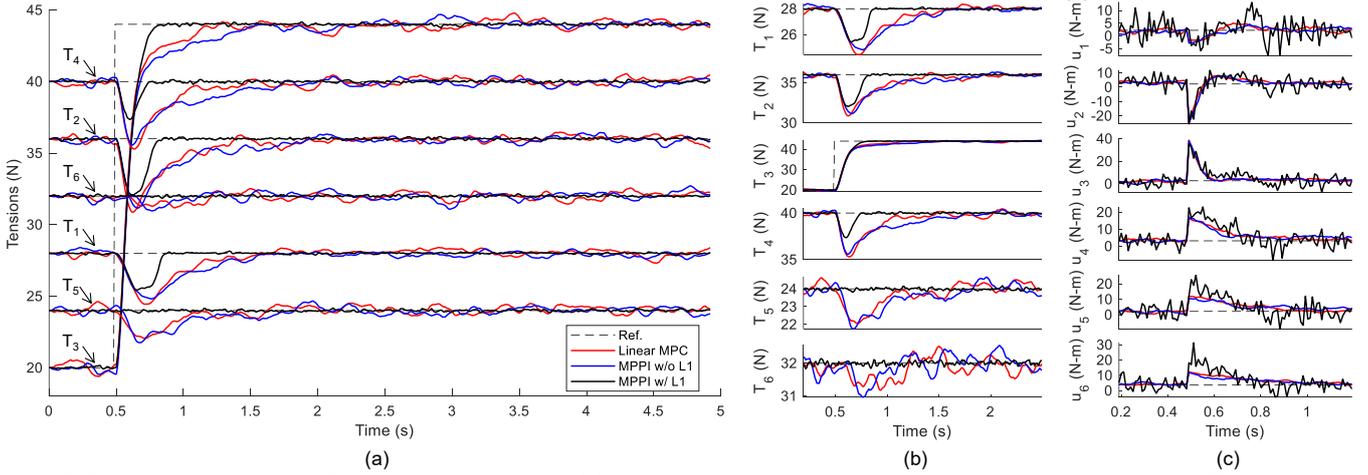

Figure 3: Tension control results. (a) The overall tension response, (b) a zoom-in on the tension response, and (c) a zoom-in on the control effort.

### 5.3 Change in velocity setpoint

This subsection tests the three controllers' ability to regulate web tension during the web speed-up phase, where the reference web speed value is increased suddenly. In this test case, all six web sections had a reference tension of 30 N. The unwinding velocity started at 0.01 m·s$^{-1}$ and then increased to 0.10 m·s$^{-1}$. The reference velocity values for the six actuated rollers all followed Eq. (15).

The results of this test case are illustrated in Figure 4. In addition, the maximum tension deviation and convergence time values are listed in Table 4. Once again, the MPPI controller without the $L_1$ term performs about the same as the LMPC controller, whereas the MPPI controller with the $L_1$ term achieves superior performance. Specifically, the MPPI controller with the $L_1$ term had an average convergence time of 0.10 s and an average maximum tension deviation of 6.01%, while the LMPC had an average convergence time of 0.66 s and an average maximum tension deviation of 9.20%. This difference once again highlights the ability of the MPPI controller to seamlessly include potentially useful non-differentiable terms in its cost function.

Table 4: Velocity setpoint change results statistics.

| Web section | | 1 | 2 | 3 | 4 | 5 | 6 |
|---|---|---|---|---|---|---|---|
| Convergence time (s) | LMPC | 0.81 | 1.00 | 0.74 | 0.63 | 0.34 | 0.42 |
| | MPPI w/o L1 | 1.21 | 1.32 | 0.99 | 1.10 | 0.60 | 0.27 |
| | MPPI w/ L1 | 0.31 | 0.30 | 0 | 0 | 0 | 0 |
| Maximum Tension Deviation (%) | LMPC | 25.1 | 12.1 | 6.97 | 5.37 | 3.33 | 2.37 |
| | MPPI w/o L1 | 25.5 | 11.7 | 9.53 | 4.60 | 4.40 | 2.50 |
| | MPPI w/ L1 | 25.5 | 7.87 | 0.63 | 0.50 | 0.67 | 0.90 |

### 6. CONCLUSIONS

This study presents an MPPI control formulation for R2R control. The MPPI approach calculates optimal control online using a Monte Carlo sampling method instead of solving an optimal control problem online like typical MPC approaches. These Monte-Carlo rollouts can be performed in parallel on a GPU, enabling online implementation even for large systems. Additionally, this approach means that MPPI control can utilize non-quadratic or even non-differentiable cost functions, which can significantly improve the performance of complex systems such as advanced manufacturing lines. To illustrate this point, a case study was performed comparing the proposed MPPI controller with a non-differentiable cost function against a standard QP-based LMPC. In the case study, the MPPI controller achieved a 70-85% reduction in average convergence time and a 35-45% reduction in average web tension maximum deviation compared to the LMPC. Thus, MPPI control is a promising method for high-precision control of large advanced R2R lines.

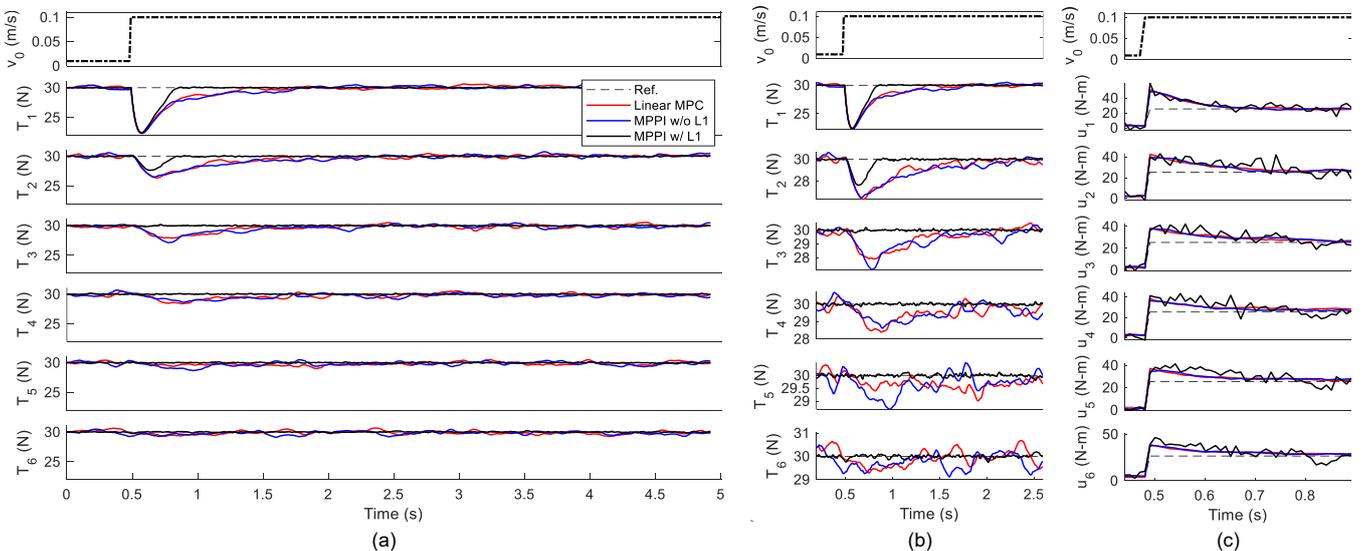

Figure 4: Velocity setpoint change results. (a) The overall tension response, (b) a zoom-in on the tension response, and (c) a zoom-in on the control effort. Each subfigure includes the unwinding velocity trajectory for reference.

## ACKNOWLEDGEMENTS

This study is based upon work supported by the National Science Foundation (NSF) under Cooperative Agreements No. EEC-1160494 and No. CMMI-2041470, as well as the NSF graduate research fellowship given to Mr. Martin. This work was also supported by the Air Force Office of Scientific Research under Grant FA9550-20-1-0101.

## REFERENCES

Bryson, A.E. (1993) *Control of spacecraft and aircraft*. Princeton university press Princeton.

Chen, Z. *et al.* (2019) 'Modeling and Register Control of the Speed-Up Phase in Roll-to-Roll Printing Systems', *IEEE Transactions on Automation Science and Engineering*, 16(3), pp. 1438–1449.

Chen, Z. *et al.* (2023) 'Robust constrained tension control for high-precision roll-to-roll processes', *ISA Transactions*, 136, pp. 651–662.

Durrett, R. (2019) *Probability: theory and examples*. Cambridge university press.

Dwivedula, R. V. and Pagilla, P.R. (2013) 'Effect of backlash on web tension in roll-to-roll manufacturing systems: Mathematical model, mitigation method and experimental evaluation', in *2013 IEEE Intern. Conf. on Control Applications (CCA)*. IEEE, pp. 1087–1092.

Gros, S. *et al.* (2020) 'From linear to nonlinear MPC: bridging the gap via the real-time iteration', *International Journal of Control*, 93(1), pp. 62–80.

Hong, N. *et al.* (2022) 'Roll-to-Roll Dry Transfer of Large-Scale Graphene', *Advanced Materials*, 34(3).

Janson, L., Schmerling, E. and Pavone, M. (2017) 'Monte Carlo motion planning for robot trajectory optimization under uncertainty', in *Robotics Research*. Cham: Springer International Publishing, pp. 343–361.

Kang, D. *et al.* (2021) 'Friction isolated rotary system for high-precision roll-to-roll manufacturing', *Precision Engineering*, 68, pp. 358–364.

Kappen, H.J. (2005) 'Path integrals and symmetry breaking for optimal control theory', *Journ. of Stat. Mech.: Theory and Experiment*, 2005(11), pp. P11011–P11011.

Martin, C. *et al.* (2025a) 'A Review of Advanced Roll-to-Roll Manufacturing: System Modeling and Control', *Journ. of Manuf. Science and Engineering*, 147(4).

Martin, C. *et al.* (2025b) 'Optimal Control of a Roll-to-Roll Dry Transfer Process With Bounded Dynamics Convexification', *Journ. of Dynamic Systems, Measurement, and Control*, 147(3), p. 031004.

Pastorelli, F. *et al.* (2016) 'The Organic Power Transistor: Roll-to-Roll Manufacture, Thermal Behavior, and Power Handling When Driving Printed Electronics', *Advanced Engineering Materials*, 18(1), pp. 51–55.

Patil, A. *et al.* (2022) 'Chance-Constrained Stochastic Optimal Control via Path Integral and Finite Difference Methods', in *2022 IEEE 61st Conference on Decision and Control (CDC)*. IEEE, pp. 3598–3604.

Patil, A. *et al.* (2023) 'Risk-Minimizing Two-Player Zero-Sum Stochastic Differential Game via Path Integral Control', in *2023 62nd IEEE Conference on Decision and Control (CDC)*. IEEE, pp. 3095–3101.

Raul, P.R. and Pagilla, P.R. (2015) 'Design and implementation of adaptive PI control schemes for web tension control in roll-to-roll (R2R) manufacturing', *ISA Transactions*, 56, pp. 276–287.

Shah, K. *et al.* (2022) 'Data-Driven Model Predictive Control for Roll-to-Roll Process Register Error', in *2022 Intern. Additive Manufacturing Conf.*. Lisbon, Portugal: ASME.

Sheats, J.R. (2002) 'Roll-to-roll manufacturing of thin film electronics', in R.L. Engelstad (ed.), p. 240.

Shelton, J.J. (1986) 'Dynamics of Web Tension Control with Velocity or Torque Control', in *1986 American Control Conference*. IEEE, pp. 1423–1427.

Todorov, E. and Li, W. (2005) 'A generalized iterative LQG method for locally-optimal feedback control of constrained nonlinear stochastic systems', in *2005 American Control Conference*. IEEE, pp. 300–306.

Wang, Y. and Boyd, S. (2010) 'Fast Model Predictive Control Using Online Optimization', *IEEE Transactions on Control Systems Technology*, 18(2), pp. 267–278.

Williams, G. *et al.* (2016) 'Aggressive driving with model predictive path integral control', in *2016 IEEE International Conference on Robotics and Automation (ICRA)*. IEEE, pp. 1433–1440.

Williams, G., Aldrich, A. and Theodorou, E.A. (2017) 'Model Predictive Path Integral Control: From Theory to Parallel Computation', *Journal of Guidance, Control, and Dynamics*, 40(2), pp. 344–357.

Xu, Y. (2009) 'Modeling and LPV control of web winding system with sinusoidal tension disturbance', in *2009 Chinese Control and Decision Conference*. IEEE, pp. 3815–3820.

Young, G.E. and Reid, K.N. (1993) 'Lateral and Longitudinal Dynamic Behavior and Control of Moving Webs', *Journal of Dynamic Systems, Measurement, and Control*, 115(2B), pp. 309–317.

Zhang, T., Tan, H. and Chen, Z. (2024) 'A Model-Based Self-Tuning Fully Decoupled Register Control Method for the Speed-Up Phase of Roll-to-Roll Systems', *IEEE Trans. on Industrial Elect.*, 71(10), pp. 12922–12930.

Zhao, Q. *et al.* (2022) 'A Dynamic System Model for Roll-to-Roll Dry Transfer of Two-Dimensional Materials and Printed Electronics', *Journal of Dynamic Systems, Measurement, and Control*, 144(7).